# History of Scientists´ Elimination of Naive Beliefs about Movement
## - The testing of the theories of Galileo in his lifetime on board of a galley -

**ABSTRACT**

Throughout the early history of Science the heliocentric world model was refused because it contradicted the thoughts of Aristotle and the medieval „Impetus" theory of movement. Even Galileo's sky observations did not lead to any acceptance of the heliocentric model, because scientists derived from Aristotle's physics that the earth was static. It was not until 1640 when Gassendi proved the principle of inertia on a moving galley, as well as Galileo's laws of free fall with a giant wheel that „impetus-physics" which incorporated naive beliefs about movement was abandoned. To understand the methods of calculating movement, novice students can be stitmulaed by using these initial experiments based on the dynamics of a ship movement and not rely on naïve approach. This can help novice students to attain a concept change from naïve to scientific conceptions about movement.

Modern students who study movement were found to harbour many of the same naive beliefs as our scientific forefathers. The history behind the elimination of these naive beliefs is presented here in order to help encourage elimination of these naive beliefs. Several studies found these naive beliefs to be held by novice students today. These naive beliefs held in the Science of movement have posed major problems for teaching novice students. Conflicting methods of modelling these movements have been the source of much controversy over time as concepts changed. Considering a conceptual change in history can be helpful in education.

**INTRODUCTION: Concept change**

In the teaching of Science usually the naive belief of students has to be changed (Chi 1981). For example, the naive belief `there is no motion without a force` should be replaced with the expert belief `there is no acceleration without a force` (diSessa 1998). Literature holds many empirical findings about eliminating naive beliefs (Chi 1994). Concept change is triggered when two concepts are in conflict and the class has to decide which concept best resolved the problem, and sees how the replacement concept explained the demonstration or quantitative problem, for example motion of a dropped object (Kalman 2004). The idea is that the evaluation of a theoretical framework does not occur until there is an alternative to produce the conceptual ecology. The specific failures such as being faced with an anomaly are an important part of this ecology (Posner 1982).

Here we demonstrate `teaching approaches in science education in the light of knowledge and understanding of the history of science` (Strike 1982) with a crucial experiment in history, which caused a concept change of scientists. This way recall of historic experiments helps students to eliminate naive beliefs. Context setting of the problem can have a significant influence on students´ reasoning and on concept change (Bao 2002). Concept change to explain the day/night cycle by the rotation of earth rather than a moving sun and a stationary earth is difficult because of students´ ontological and epistemological presuppositions (Vosniadou 1992). In order to achieve conceptual change, the presuppositions have to be subjected to experimentation and falsification (Vosniadou 1994). For example the concept of inertia is tested by the force concept inventory (a ball leaving a circular path), only 5% of all students answered correctly (the path tangential to the circle) independent of distracters in the multiple-choice answers (Rebello 2004, question 7). There are similar results concerning a ball falling from a moving airplane (Rebello 2004, question 14). This is explained by diSessa (1993) by phenomenological primitives used by students (p-prims, diSessa 1998). Since ´both learning and the history of science requires us to understand how conceptions change´ (Strike 1982), we investigate both the concept change from Medieval to Newtonian mechanics as well as the first statement of principle of inertia in this article.

**DATA COLLECTION. Elimination of naive beliefs about movement in history**

A study is conducted here in which the history behind the elimination of common naive beliefs about movement by scientists is presented in order to help encourage it (Monk 1998). It is demonstrated finally that these naive beliefs are still found to be held by novice students today. Archimedes of Syracuse mentioned in his script „The Calculator of Sands" that an early Greek astronomer named Aristarch explained the movements in the sky by a theory that the earth circulates around the sun like the other planets do. But no one believed that, because the earth seemed to stand still  (Ley 1963). The same happened when Copernicus, who gave his name to the modern heliocentric system, claimed the same thing (Ley 1963). In the time of Galileo science still thought about the world in Aristotle's terms of dichotomy meaning that the planet earth is made up of the elements water, air, fire and earth while the sky, the other planets as well as the stars, are made of aether. Dichotomy was seriously shaken by Galileo's observations of the sky by telescope laid down in his famous script „Sidereus Nuncius" (News from New Stars 1610 / 1611, Galileo 1967). However, the scientific world did not accept these findings as proof of the heliocentric system, because science was used to a redesigned kind of physics of Aristotle's called the „theory of Impetus"(Ibn Sinha 1885/86). This theory claimed the necessity of a `continuous mover` as a cause of every movement (Abu'l Barakat 1939), in modern algebraic language „velocity is a force divided by resistance", this idea was elaborated on by many a scientist such as Avicenna (965-1039) and Albert of Saxony (1316-1390). However, scientists of antiquity as well as Arabian and Renaissance became aware of contradictions to observations as well as other antiquity theories such as the atomic theory of Epicures. Even simple observations like the trajectory of an arrow caused enormous difficulties within the theory of Impetus. Epicures (341-270 B.C.) described atoms as not infinitely small, invisibly small and continuously in movement. This theory of atoms contradicts Aristotle's physics and the impetus theory. Consequently, atomic theory was abandoned in medieval times as well as in $16^{th}$ century physics (Sambursky 1974).

In modern physics force is the cause of the change of movement (acceleration), contrarily, impetus calls for a cause for movement at any time based on Aristotle's claims for rest as the privileged and natural state on earth (Brown 2006). Based on these ideas, Aristotle's, Ptolemaist, and Thycho Brahe, amongst many, claimed that, provided the earth rotates in an easterly direction, a stone thrown up would come down in the west as long as earth moved in an easterly direction during the time of the throw, and birds and clouds would drift rapidly in an westerly direction. Since this cannot be observed, it should be reasoned that earth stays still and does not move (Sambursky 1974). Arguing that way the heliocentric ideas of Aristarch from Samos, Copernicus and Galileo have all been rejected. Even Galileo's revolutionary observations of mountains on the moon, of spots on the sun, phases of Venus and four moons encircling Jupiter made by a telescope in 1609 and in 1610 all contradicting the antiquity system of Ptolemaist did not change the situation (Galileo 1957). The seeming stillness of the earth was the major point in both the cases in the courts of inquisition against Galileo in 1516 and 1533, which led to the denouncement of the „Dialogue concerning the chief world systems" by Galileo.

Galileo used the principle of inertia in his arguments in the „Dialogue". On the second day of the „Dialogue" Salviati, who is Galileo's spokesman of the heliocentric system, claims a stone falling down the mast of a moving ship will hit the ground at the foot of the mast and will not drift to the stern as was claimed by Aristotle. Simplicio, Aristotle's spokesman of the Ptolemaic system, does not accept this because that assertion is a claim and has never been proven (Galileo 1967). This dialogue resembles teaching physics to novice students holding naive beliefs about movement. Such an experiment with a moving ship would indeed have been too elaborate for Galileo. In this very situation this crucial experiment with the ship became reality with the help of the French priest and scientist Gassendi.

This experiment can help eliminate naive beliefs held by novice students today.

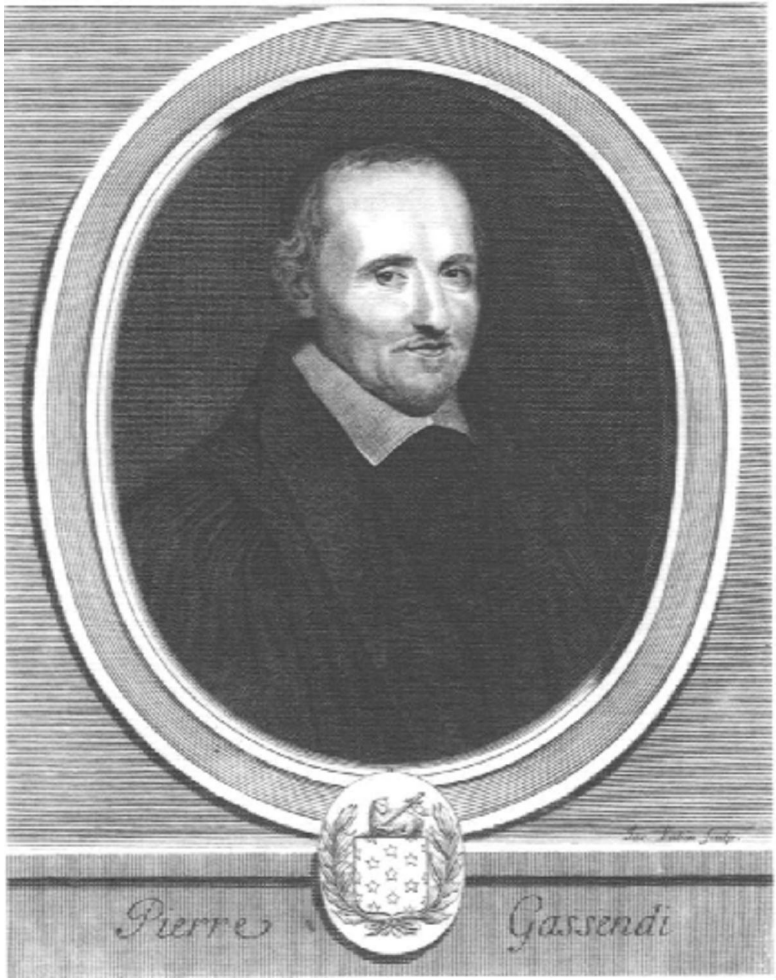

Fig. 1 Portrait of Pierre Gassendi 1592-1655, from: Egan , 1964

Pierre Gassendi was born in 1592 in Champtercier in Provence; he studied theology in Aix-en-Provence and Digne (Jones 1981) was consecrated as a priest (minister) in Digne and became a professor in Aix (Detel 1978, Egan 1964). Travelling frequently to Paris he co-operated with the scientific circle (salon) of Mersenne. Since scientific academies had not been founded, this circle of Mersenne was the most important scientific audience at that time.

The discussions there caused Gassendi to perform several experiments, such as:

- He repeated Pascal's measurement of ambient air pressure, also producing a vacuum renouncing Aristotle's theory of „horror vacui"
- He observed Mercury's eclipse of sun by telescope in 1630
- He was the first to draw a map of the moon to scale; a crater was named „Gassendi" in memory of this,
- He observed the growth as well as the solution of crystals through a microscope deriving existence of atoms from the persistence of angles in crystals (Detel 1978, Fisher 2005).
- He discovered of the rings of Saturn before Huygens did so, describing Saturn's view through the telescope as elongated and wrinkled with handles.

He had exchanged letters with Galileo since 1625 (Detel 1978). After his death in Paris in 1655 his work about logic was printed in Tours in 1658 (Jones 1981), mentioning his experiments on a ship.

**DATA COLLECTION AND ANALYSIS.  Reconstruction of the experiments aboard a galley 1640 which proved the principle of inertia and eliminated naive beliefs about motion**

An understanding of the history of Science regarding motion can promote student understanding of important concepts through the process of conceptual change. The argument is that there are similarities between student naive beliefs and the early naive beliefs of motion held by scientists. Gassendi´s straightforward ship experiments demonstrated the effectiveness of inertial explanation of motion. The students themselves in an appropriate conceptual change teaching environment could perform similar experiments. Before recapping these historical events it should be noted that in 1610 Galileo published *Sidereus Nuntius* with discoveries of the moons of Jupiter, mountains on the moon and suns´ spots. In 1616 the inquisition forced him to mention the heliocentric system of Copernicus only as a hypothesis rather than a fact. Galileo's´ *Dialogue Concerning the Two Chief World Systems comparing* the heliocentric system of Copernicus with the Ptolemaic earth centred world system in a dramatic dialogue was published in Italian in 1632 but censored by the Roman Inquisition in 1633. The translation into Latin was published in 1636 in Strasbourg in protestant oriented central Europe, ignoring censorship. The 2$^{nd}$ Latin edition was improved by Galileo in 1641; it became his last major work and was released all over Europe (Galileo 1957, 1967). Since Galileo exchanged letters with Gassendi (Taussig 2004) it can be conjectured that Galileo knew the ship experiments when he was working on the 2$^{nd}$ edition.

Several ship experiments performed by Pierre Gassendi in 1640 decided whether Galileo's theoretical experiment on the movement of bodies in a cabin of a ship was correct. This question was evidently answered by Gassendi positively. This episode documents the importance of access to proper experimental equipment in the testing of mechanical hypotheses, and of the necessity of political support in the present case, in which expensive military hardware was required.

The friend and sponsor of Gassendi Louis Emanuel of House Valois became governor of Provence in 1638 and so he was in charge of the French fleet. Thus it became possible to conduct the experiment with a moving ship as proposed by Galileo. This experiment was described in his letter „De motu impresso epistolae due", Paris, 1640 / 1642 (Gassendi 1642) in articles V and VIII superficially. Data collected from the Bavarian State Library as well as the „Musée de la Marine" in Paris was analysed to give a report of the valid construction of Gassendi´s experiments to prove Galileo's theories. By these means the experiments can be reconstructed as follows. Ships sailing are exposed to waves and heel over and thus are not suited to perform these experiments, small rowing boats have no masts. For testing Aristotle's claims regarding motion large military galleys are required to which Galileo had no access. The experiments could be done with Louis Emanuel of House Valois and his access to the French fleet at Marseilles. The galleys of the French fleet at 1640 had a length of about 40m, the masts were about 20 m high, and attained the speed of modern sports rowing boats, about 5m/s (Mondfeld 1972). At this time Galleons of the French fleet presented two masts as shown in figures 2 and 3.

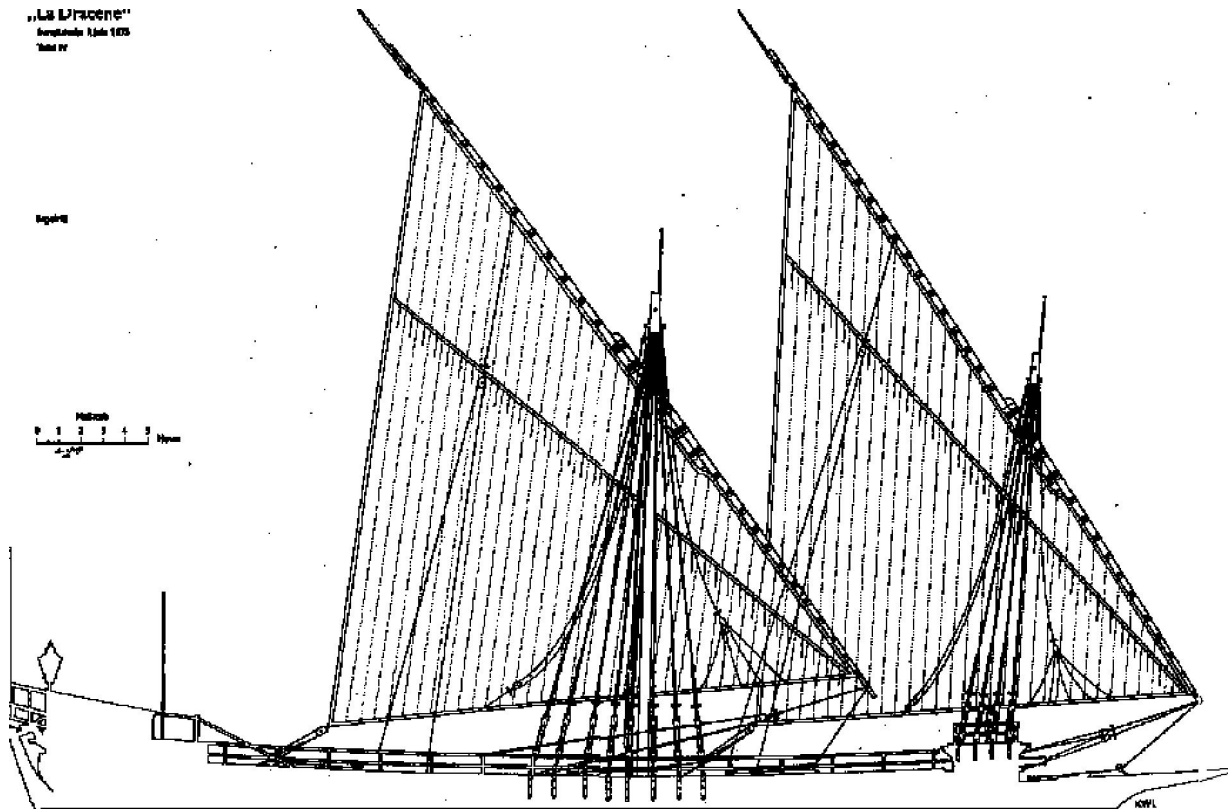

Fig. 2  French Galley in 1675, „Musée de la Marine", Paris, from: Mondfeld  1972

The heights of the mast were approximately 25 m of which 20 m were accessible, thus the free fall from the mast lasted approximately 2 seconds while the ship was swiftly rowed a distance of about 10m. Thus the falling stone would be carried away 10m towards the stern. Gassendi wrote in his letter, however, that the impact of the stone was right at the foot of the mast - a very impressive contradiction to the movement theories of Aristotle as well as "impetus" and thereby a major defeat for the entire medieval scholar system of „Scholasticism". Today, discussion of this experiment can be a way to help students to eliminate naive belief  about movement.

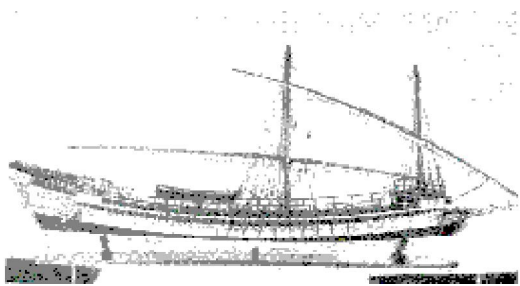

Fig. 3  Galley of French fleet 1692, „Musèe de la Marine", Paris , from: Mondfeld 1972

Gassendi performed other experiments aboard the moving ship. He let bowls down the deck - once from the stern and once from the bow of the ship rolling one time in and the other time against the movement of the vessel. They arrived at the same time. The galleys of the French fleet in 1640 had 1- to 36 - pound cannons aboard. The deck was elevated at both ends (stern and bow) about 0.5m with respect to amidships, along a length of 17 m respectively (Mondfeld 1972, Risch 2007). If you let a cannonball roll from either side, it would attained the speed of about 2.5 m/s taking into account the energy of rotation, this is less than the speed of the boat of about 5m/s. If Aristotle's theories about movement and the "impetus" theory were right, then the cannonball released at the stern would not even move toward amidships, but move away towards the stern. This again was an impressive contradiction to the Aristotle and "impetus" theories of movement and thereby a major defeat for all medieval physics as well as a demonstration today helping novice students to eliminate naive beliefs.

Having completed experiments aboard the moving ship, Gassendi has written the principle of inertia for the first time as it is used today in article XVI of his first „de motu" letter: „You will ask in passing what would happen to that stone which I claimed could be imagined in empty space if it were roused from its state of rest and impelled by some force. I answer that it is probable that it will move indefinitely in a uniform fashion, slowly or rapidly, depending on whether a small or great impetus had been imparted on it. I take my proof from the uniformity of the horizontal motion I have already explained since it would apparently not stop for any other reason than then influence of perpendicular motion... would not be accelerated or slowed down, and therefore would never stop." (Detel 1978, Fisher 2005). The first two „de motu" letters were addressed to Pierre Dupui in Paris in November and December 1640 and were published in 1642 (Brush 1972). Galileo's name is used today with the principle of inertia; he claimed this principle before for rotary motions alone (Galileo 1967, Brown 2006).

High expenses for the ship's experiments with French state galleys and more than one hundred participants including persons of importance gave these refusals of Aristotelian physics and scholastic teachings credibility and public focus far beyond the borders of France (Taussig 2004). Due to expenses, the number of participants and high prestige of galleys, which had been involved in state races (Rambert 1931); this can be called the first „big Science" experiment. This paved the way for wide acceptance of Galileo's findings and the heliocentric model as well as scientific revolution. These experiments can be used to eliminate naive beliefs about movement found to be held by novice students.

## METHODOLOGY. Free fall experiments to eliminate naive beliefs about movement

The „de mute" letters took away the strongest objection against the heliocentric model by disproving Aristotelian physics. However, rejection of the findings of Galileo did not cease. Especially the „Dialogue Concerning the Two Chief World Systems", and his experiments about free fall were still rejected.  The naive beliefs about free fall kept at those times are found to be held by novice students today (Bar 1994). With his inclined plane (shown in figure 4), Galileo proved the law of odd numbers, claiming distances passed in consecutive times behave like odd numbers; in today's words, speed increases proportionally with time.

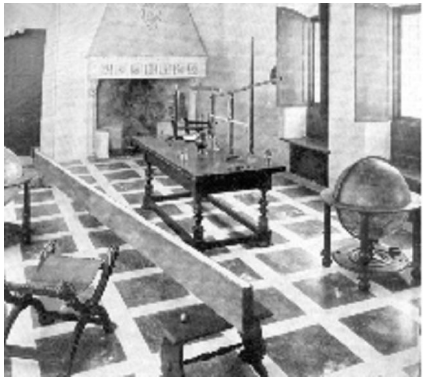

Fig. 4 Galileo's inclined plane, reconstruction in the Deutsche Museum, Munich, Germany.

This is in contrast to medieval scholastic physics claiming speed would increase in equal steps with distance deriving this from the writings of Aristotle. This would yield to an exponential increase of distance with time. Having no modern clock, Galileo probably used his heartbeat or a water clock as a measure of time (Crawford 1996). Such poor accuracy could not clearly distinguish scholastic teaching from Galileo's law of free fall. Galileo himself describes precisely a comparison of time intervals of free fall in his „Dialogue Concerning the Two Chief World Systems": He drops ammunition for a musket as well as a cannonball from the inclined tower of Pisa, both hit at ground the same time (Galileo 1967). It is not clear if he really performed this experiment.

For a final proof of Galileo's law of free-fall, Gassed replaces measurement of time by time difference in his shrewd wheel of free fall (figure 5). This treadmill wheel with almost 4m diameter contained three glass tubes, which guided three bodies falling simultaneously. They arrived the same time proving Galileo's law of free fall.

Since the wheel represents Thales´ circle, force accelerating the bodies as well as distance are diminished by the same factor (cosine) resulting in the same time of fall, provided Galileo's law holds. The time would not be the same if speed would increase proportionally with distance, as was claimed by scholastic theory.

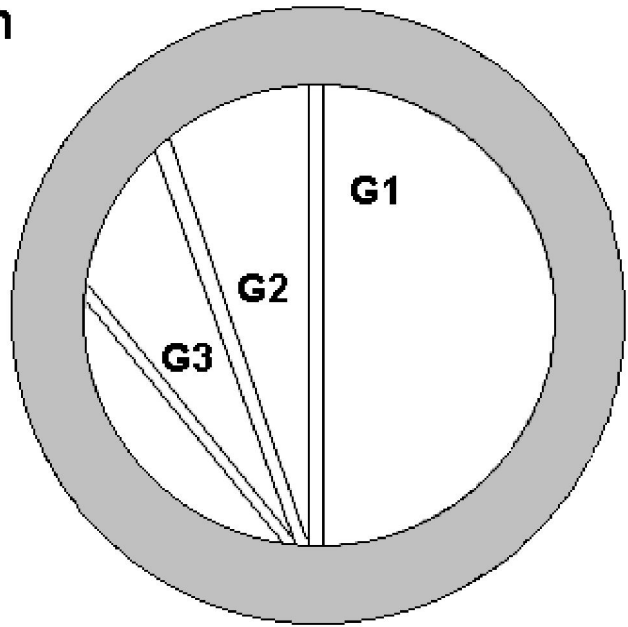

Fig. 5, Gassendi´s wheel of free fall with three glass tubes G1 to G3

This experiment could be repeated everywhere improving acceptance of Galileo's laws of free fall. In our times it can still serve as an impressive demonstration for students. The simultaneous arrival of three falling bodies is more convincing to novice students than digital readouts of computers or counters as it is used in many present day classroom experiments about free fall.

Discussing free fall in his „de mutu" letter, Gassendi draws a parabolic curve of throw claiming that speed is diminished as a body is thrown up the same way as it increases falling down following Galileo's law of odd numbers ("reciproce", „de motu" article VII).

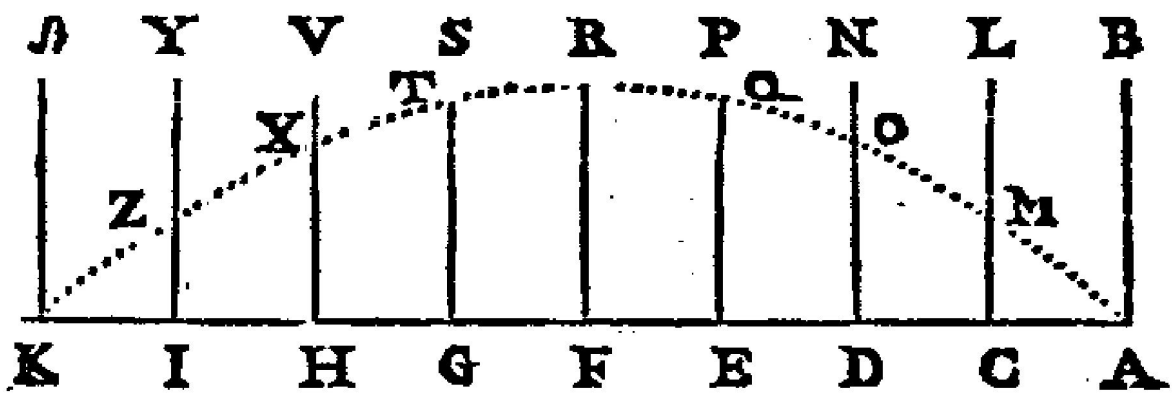

Fig. 6. Parabolic path of body thrown, possibly the first drawing of physics in a modern coordinate system, from Gassendi, 1642.

This graphic representation (Figure 6) of Galileo's law of odd numbers is possibly the first visualisation of a law of physics in a modern co-ordinate system. His arguments, which reciprocate Galileo's law of diminishing odd numbers when a body is thrown up, imply invariance of physics laws with respect to time variance.

After his death the „Institutio Logica" (Jones 1981) was published in Tours trying to simplify Aristotle's logic by a new scheme. The fourth part concerns methods of logic arguing and in its fourth canon he stresses logical implications of his experiments on a moving ship. Canon four conjectures that both reason and senses have to be taken into account arguing logically. The senses, however, have to be trusted more than reason, because „there is the possibility that the reasoning is an inaccurate estimate or surface explanation only, the true reason for the phenomenon appearing to the senses as it does remaining hidden. For instance, although reason would at first persuade us that an arrow fired from the stern of a moving ship would fall not upon the same stern but into the sea some distance away, the ship having moved forward in the meantime, yet reason must give way to the senses, because experience shows us that it turns out differently and in actuality motion is imparted to the arrow not only by the bow but also by the ship itself (Jones 1981)." This example contradicts Aristotle. It is also a description of conceptual background naive beliefs held by many novice students today. The next example of Gassendi´s is based on Aristotle: „...People who used to believe that there were no people living in the Antipodes did so on the grounds that if there were they would fall downwards into the sky, but now that they have been discovered to indeed exist, this reasoning obviously loses its validity in the light of experience..."

**DISCUSSION.  Consequences from Gassendi´s experiments to teaching physics**

Gassendi´s experiments are important for modern psychology of perception and also for teaching. Teaching movement of earth is still a difficult task today (Parker 1998).  In many investigations about perception of movement, people believed that bodies lose their speed instantaneously and stop as soon as the force driving ceases. The psychologist McCloskey found that most students believe a ball dropped while running will fall down not in a parabolic fashion but abruptly and straightaway to ground (McCloskey 1980, Caramazza 1981). This phenomenon is called „intuitive physics" (McCloskey 1980, Krist 1993). Similarly, Clement (1982) found that most students believe a rocket drifting in space will retain its speed before ignition (figure 7) when engines stop.

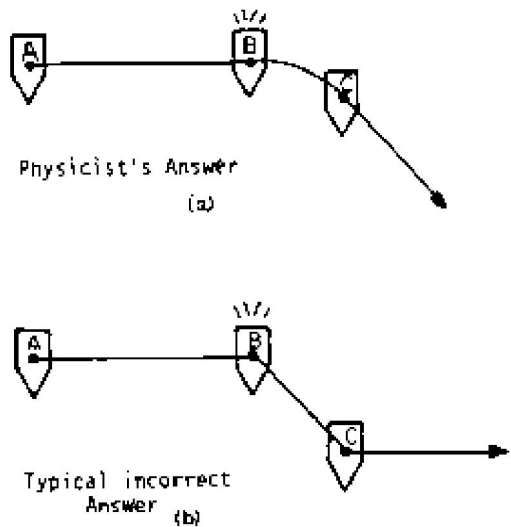

Fig. 7  Rocket path question to investigate intuitive physics.

College students were asked about an object dropped from a mast of a moving ship, most students answered the object will fall and strike the deck behind the foot of the mast rather than at the foot of the mast (Whitaker 1983). Though „intuitive physics" ignores the principle of inertia and follows Aristotelian physics and the false idea of impetus, it has been found in many investigations about students (Reif 1987, Clement 1989, Adey 1992, DiStefano 1996a, 1996b, Gautreau 1997). It is a serious pitfall when teaching beginner's classes in physics, therefore testing scores of students in physics remain low (Driver 1989, Licht 1990).

**CONCLUSION. Possible ways to improve concept change in teaching and learning.**

Beginner's difficulties in physics can be eased by the discussion of Gassendis´ experiments and repetition of these by students' observing for example falling bodies in a moving vehicle such as a car, a bus, a lorry or a boat as an out- of- school experience (Mahoy 1997). This kind of experience has been shown to improve learning (Donnelly 1998).
To ease elimination of naive beliefs about movement, each individual can recall the long way mankind took to overcome these naive beliefs.

Even a single experiment can induce a concept change in students; most efficient are hands-on experiments (Abbott 2000). Individual naive belief in physics is sometimes a recapitulation of a historic naive belief (Chi 1994). Approach to concept change is easier `in the light and understanding of the history of science` (Strike 1982). For example, computer simulation of Galileo's historic experiments with the inclined plane and water clock helped understanding of physics concepts (Borghi 1992). Suspense stories like "*Galileo and the inquisition*" raise emotions improving memory and attendance thus easing concept change, according to results of psychological research (Cahill 1994, Erk 2003).

FIGURE CAPTIONS

Fig.  1 Portrait of Pierre Gassendi 1592-1655

Fig. 2  French Galley in 1675, „Musèe de la Marine", Paris, (Mondfeld  1972)

Fig. 3  Galley of French fleet 1692, „Musèe de la Marine", Paris (Mondfeld  1972)

Fig. 4  Galileo's inclined plane, reconstruction in the Deutsche Museum, Munich, Germany.

Fig.  5, Gassendi´s wheel of free fall with three glass tubes G1 to G3

Fig.  6. Parabolic path of body thrown, possibly the first drawing of physics in a modern co-ordinate system

Fig. 7  Rocket path question to investigate intuitive physics.